\documentclass[prd,nofootinbib,showpacs]{revtex4}

\usepackage{graphicx}% Include figure files
\usepackage{dcolumn}% Align table columns on decimal point  
\usepackage{bm}% bold math
\usepackage{mathrsfs,amsmath}
\usepackage{hyperref}
\usepackage{color}

\newcommand{\beq}{\begin{equation}}
\newcommand{\eeq}{\end{equation}}

\begin{document}

\title{CosmicFish Validation Notes V1.0}

\author{Marco Raveri$^{1,2,3}$, Matteo Martinelli$^{4,5}$, Gong-Bo Zhao$^{6,7}$ and Yuting Wang$^{6,7}$}
\affiliation{
\smallskip
$^{1}$ SISSA - International School for Advanced Studies, Via Bonomea 265, 34136, Trieste, Italy \\
\smallskip
$^{2}$ INFN, Sezione di Trieste, Via Valerio 2, I-34127 Trieste, Italy \\
\smallskip
$^{3}$ INAF-Osservatorio Astronomico di Trieste, Via G.B. Tiepolo 11, I-34131 Trieste, Italy \\
\smallskip
$^{4}$ Institute Lorentz, Leiden University, PO Box 9506, Leiden 2300 RA, The Netherlands \\
\smallskip
$^{5}$ Institut f\"ur Theoretische Physik, Ruprecht-Karls-Universit\"at Heidelberg, Philosophenweg 16, 69120 Heidelberg, Germany.  \\
\smallskip
$^{6}$ National Astronomy Observatories, Chinese Academy of Science, Beijing, 100012, P.R.China \\
\smallskip
$^{7}$ Institute of Cosmology \& Gravitation, University of Portsmouth, Dennis Sciama Building, Portsmouth, PO1 3FX, UK
}

\begin{abstract}
These notes show and comment the examples that have been used to validate the CosmicFish code.
We compare the results obtained with the code to several other results available in literature finding an overall good level of agreement.
We will update this set of notes when relevant modifications to the CosmicFish code will be released or other validation examples are worked out. \\
The CosmicFish code and the package to produce all the validation results presented here are publicly available at~\url{http://cosmicfish.github.io}. \\
The present version is based on CosmicFish Jun16.
\end{abstract}

\date{\today}


\maketitle

\tableofcontents
\section{Introduction}
In~\cite{Raveri:2016xof, Raveri:2016leq} we introduced the CosmicFish code as a powerful tool to perform forecast on many different models with future cosmological experiments. \\
In this set of notes we show the validation pipeline that was used for the code. We compared the results obtained with the CosmicFish code to other results in literature. We find an overall good level of agreement. \\
Together with these notes we release a CosmicFish package that contains the relevant code to produce all the results presented here. This package is going to be updated as new validation results become available.
The CosmicFish code and its validation package are publicly available at~\url{http://cosmicfish.github.io}.

\section{CMB Forecasts}
\subsection{Planck Blue Book and CMBpol}
\begin{figure}[h!]
\begin{center}
\begin{tabular}{cc}
\includegraphics[width=8cm]{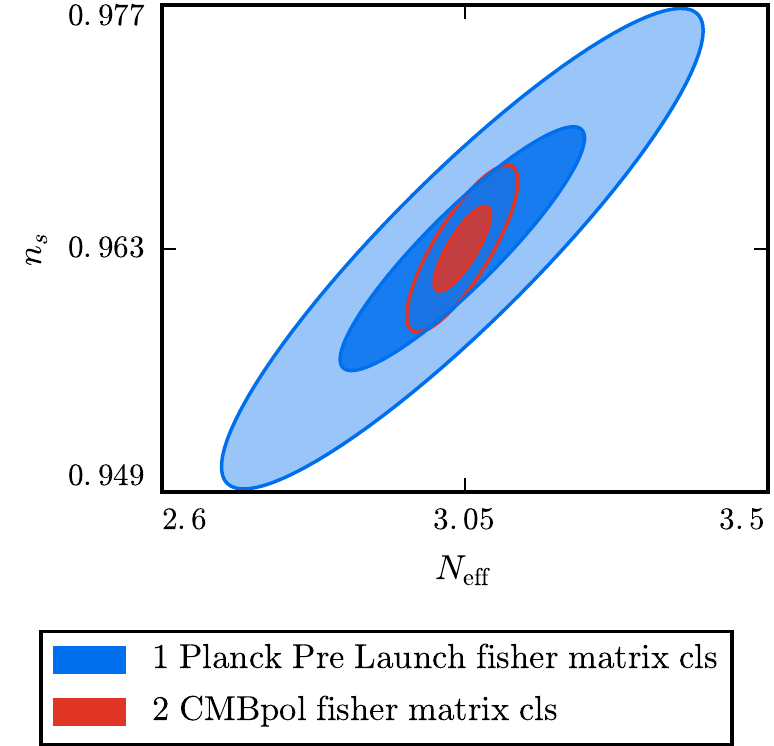}&
\includegraphics[width=8cm]{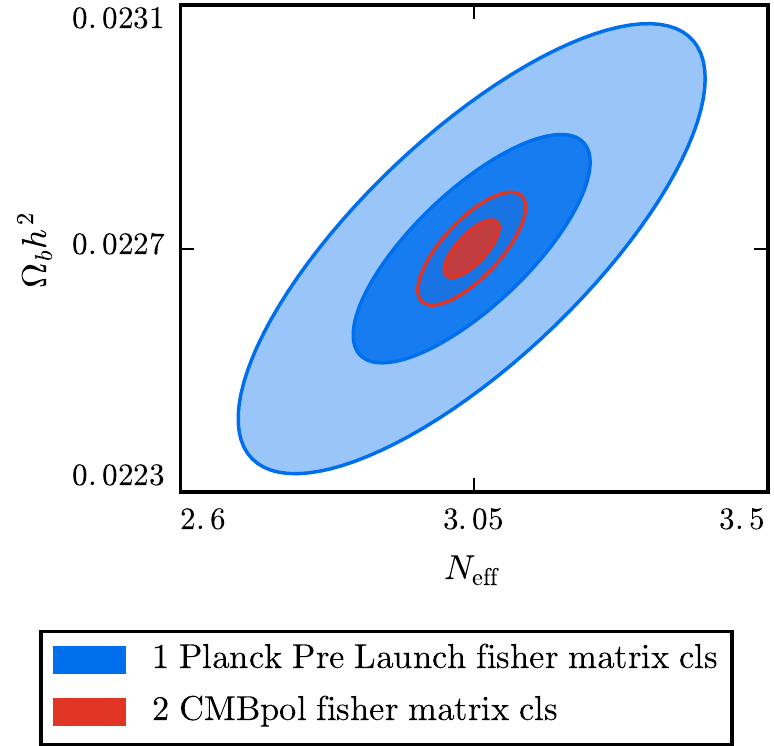}\\
\includegraphics[width=8cm]{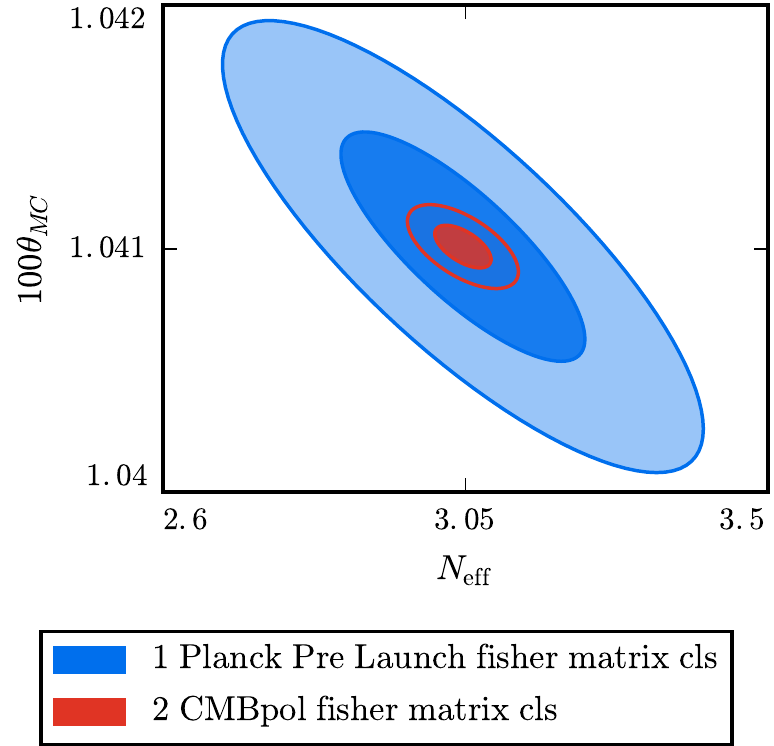}&
\includegraphics[width=8cm]{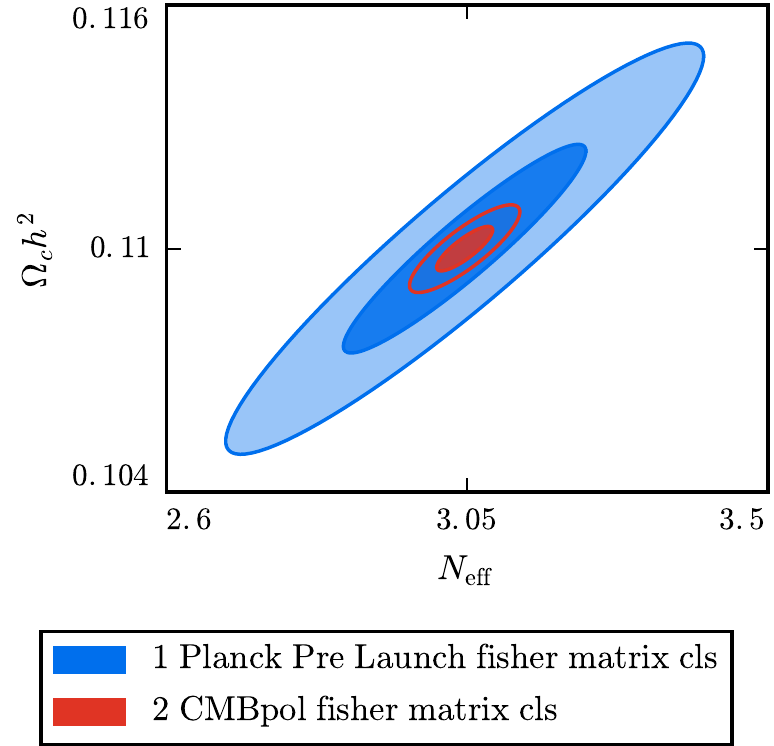}\\
\end{tabular}
\caption{Contour plots between $N_{eff}$ and standard parameters. Moving clockwise from top left panel, these
can be compared with Figures 3, 4, 5 and 6 of~\cite{Galli:2010it}. Blue and red contours here 
correspond to blue and green contours on the paper.}\label{fig:cmbfor}
\end{center}
\end{figure}

The validation package contains forecasted results for the {\it Planck} mission, obtained using {\it Planck} Blue Book specifications~\cite{bluebook}, and for the proposed CMBpol satellite~\cite{2009arXiv0906.1188B}.\\
CosmicFish is used varying the standard 6 parameters ($\Omega_bh^2,\ \Omega_ch^2,\ \tau,\ n_s,\ \log{(10^{10}A_s)},\ h$), from now on dubbed as S6, and the number of relativistic species $N_{eff}$.\\
The results are compared with what is obtained in Section III.C of~\cite{Galli:2010it}; even though these results were  obtained with MCMC methods, CosmicFish bounds and degeneracies show a very good agreement (see Figure~\ref{fig:cmbfor} and Table~\ref{tab:cmbfor}).\\

\begin{table}[h!]
\begin{center}
\begin{tabular}{|c|c|c|c|}
\hline
\hline
Parameter          & Fiducial & {\it Planck} $68\%$ c.l. bound & CMBpol $68\%$ c.l. bound \\
\hline
$\Omega_bh^2$      & $0.0227$ & $0.0002$ & $5\times10^{-5}$\\
$\Omega_ch^2$      & $0.11$   & $0.003$  & $0.0005$\\
$100\theta_{MC}$   & $1.041$  & $0.0005$ & $9\times10^{-5}$\\
$\tau$             & $0.09$   & $0.004$  & $0.003$\\
$n_s$              & $0.963$  & $0.007$  & $0.002$\\
$\log{10^{10}A_s}$ & $3.18$   & $0.01$   & $0.005$\\ 
$N_{eff}$          & $3.0$    & $0.2$    & $0.04$\\          
\hline
\end{tabular}
\caption{$68\%$ confidence level bounds on cosmological parameters obtained through {\it Planck} and CMBpol
forecasts. This can be compared with Table IV of~\cite{Galli:2010it}.}\label{tab:cmbfor}
\end{center}
\end{table}
\newpage
\subsection{Planck 2015}
\begin{figure}[h!]
\begin{center}
\begin{tabular}{cc}
\includegraphics[width=14cm]{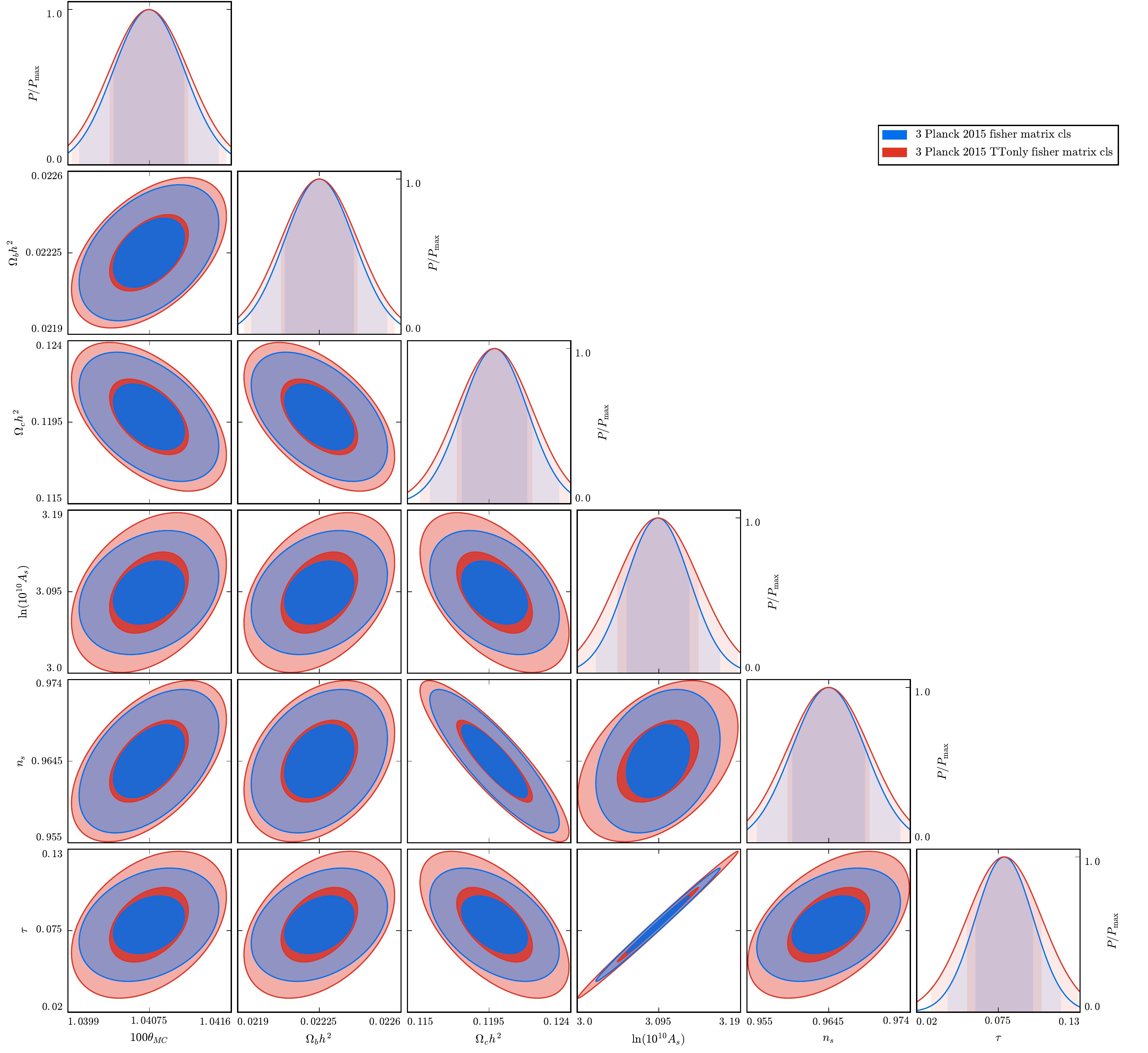}
\end{tabular}
\caption{Contour plots for standard cosmological parameters obtained with {\it Planck} 2015 $TT$ only (red contours) and with the combination $TT+TE+EE$ (blue contours), to be compared respectively
with the red and blue contours of Figure 6 of~\cite{Ade:2015xua}.}
\label{fig:pk}
\end{center}
\end{figure}

The CMB pipeline is also validated using {\it Planck} 2015 real performances specifications~\cite{Adam:2015rua}, which allow to produce bounds on the cosmological parameters mimicking the performances of the real experiment.
Figure~\ref{fig:pk} and Table~\ref{tab:pk} show the results obtained varying the S6 parameters both using only temperature spectra and including also $EE$ and $TE$.\\
A comparison of the $TT$ results with~\cite{Ade:2015xua} highlights good agreement with the {\it Planck} 2015 results, with the exception of the $\tau$ and $A_s$ parameters, due to the fact 
that in our analysis the $lowP$ {\it Planck} polarization at small multipoles is not included. More complicated is the comparison when the polarization spectra are considered; 
the {\it Planck} likelihood analysis relies on a modelization of foreground effects based on some nuisance parameters, which is not yet included in CosmicFish. In order to partially mimic the 
effect of these parameters on the constraining power brought by CMB polarization, we strongly reduce the sky fraction $f_{sky}$ observed for polarization to $0.01$. 
The bound obtained this way are compatible with {\it Planck} 2015 results.
\begin{table}[h!]
\begin{center}
\begin{tabular}{|c|c|c|c|}
\hline
\hline
Parameter            & Fiducial & $TT$ $68\%$ c.l. bound & $TT+TE+EE$ $68\%$ c.l. bound\\
\hline
$\Omega_bh^2$        & $0.0222$ &    $0.0002$            & $0.0001$\\
$\Omega_ch^2$        & $0.12$   &    $0.002$             & $0.002$\\
$\theta$             & $1.0407$ &    $0.0004$            & $0.0004$\\
$\tau$               & $0.08$   &    $0.03$              & $0.02$\\
$\log{(10^{10}A_s)}$ & $3.09$   &    $0.05$              & $0.04$\\
$n_s$                & $0.964$  &    $0.005$             & $0.004$\\ 
$h$                  & $0.673$  &    $0.009$             & $0.008$\\
$\Omega_m$           & $0.32$   &    $0.01$              & $0.01$\\
$sigma_8$            & $0.83$   &    $0.02$              & $0.01$\\
\hline
\end{tabular}
\caption{$68\%$ confidence level bounds on cosmological parameters obtained using {\it Planck} 2015 $TT$ only and {\it Planck} 2015 $TT+TE+EE$ to be compared with the second and fifth column
of Table 3 of~\cite{Ade:2015xua}.}\label{tab:pk}
\end{center}
\end{table}

\newpage
\section{Redshift Drift Forecasts}
\begin{figure}[h!]
\begin{center}
\begin{tabular}{cc}
\includegraphics[width=8cm]{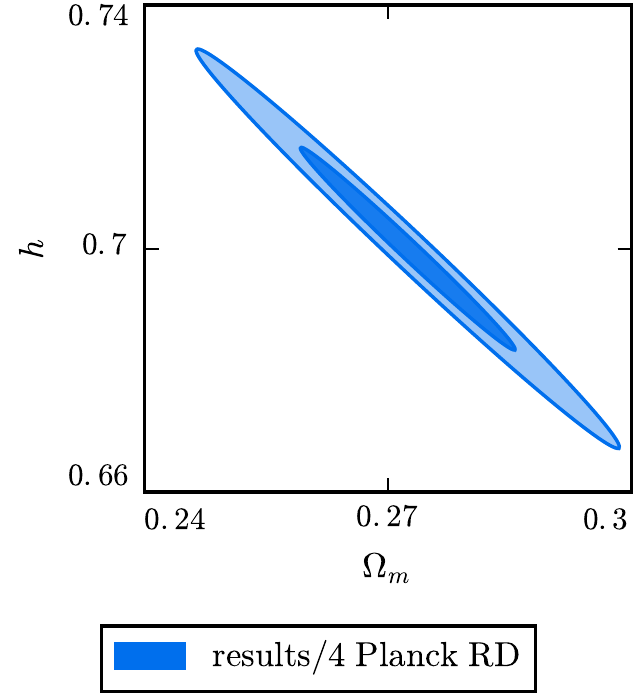}&
\includegraphics[width=8cm]{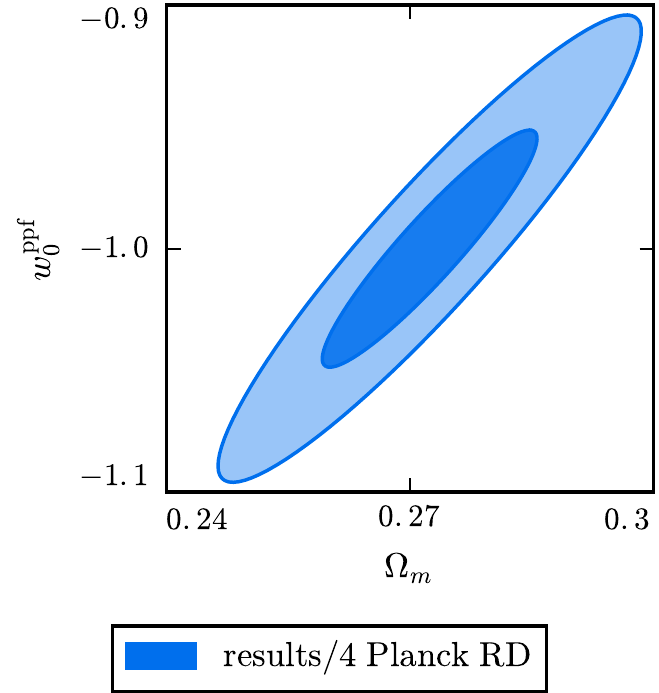}\\
\end{tabular}
\caption{Contour plots between $\Omega_m$ and $h$ (left panel) and between $\Omega_m$ and $w_0$. 
These plots can be compared respectively with the green contours of Figures 5 and 4 of~\cite{Martinelli:2012vq}.}
\label{fig:rd}
\end{center}
\end{figure}

CosmicFish includes a Fisher matrix forecast module for redshift drift.
This observable, considered alone, is not strongly constraining so we expect results to be biased by non-Gaussian features in the likelihood.
We therefore validate this observables only in combination with CMB forecast. \\
As of redshift drift observations we consider E-ELT specifications as used in~\cite{Martinelli:2012vq}\footnote{notice 
these are not the most up to date specifications for E-ELT redshift drift measurements; they are used only for 
validation purposes.}, while for CMB we use {\it Planck} Blue Book specifications.\\
In this case CosmicFish is used with S6 parameters to which the possibility of a constant $w_0$ different from
$-1$ is added. The results are then compared with Section V of~\cite{Martinelli:2012vq}, finding a good agreement
with the MCMC results obtained there (see Figure~\ref{fig:rd} and Table~\ref{tab:rd}).
Notice that bounds are slightly looser in our analysis; this is due to the inclusion of an HST prior in the analysis of~\cite{Martinelli:2012vq}.

\begin{table}[h!]
\begin{center}
\begin{tabular}{|c|c|c|}
\hline
\hline
Parameter          & Fiducial & {\it Planck}+E-ELT $68\%$ c.l. bound \\
\hline
$\Omega_bh^2$      & $0.0226$ & $0.0001$\\
$\Omega_ch^2$      & $0.1109$ & $0.0009$\\
$100\theta_{MC}$   & $1.0397$ & $0.0003$\\
$h$                & $0.71$   & $0.01$\\
$\Omega_m$         & $0.27$   & $0.01$\\
$w_0$              & $-1.0$   & $0.05$\\        
\hline
\end{tabular}
\caption{$68\%$ confidence level bounds on cosmological parameters obtained through {\it Planck}+E-ELT
forecasts. This can be compared with Table II of~\cite{Martinelli:2012vq}.}\label{tab:rd}
\end{center}
\end{table}
\newpage
\section{Supernovae Forecasts}
\begin{figure}[h!]
\begin{center}
\begin{tabular}{cc}
\includegraphics[width=8cm]{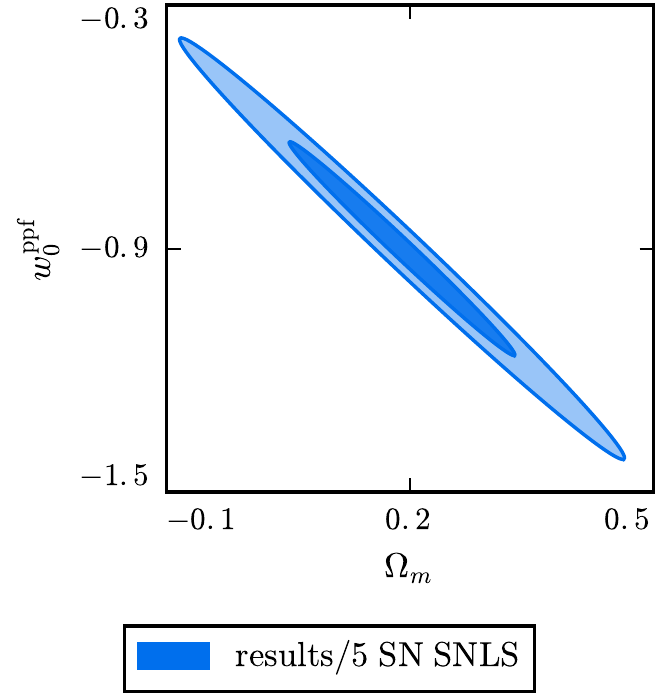}
\end{tabular}
\caption{Contour plot between $\Omega_m$ and $w_0$ to be compared with Figure 4 of~\cite{Conley:2011ku}.}
\label{fig:sn}
\end{center}
\end{figure}

The CosmicFish Supernovae pipeline is validated using as free parameters only the constant Dark Energy equation
of state parameter $w_0$ and the baryon and cold dark matter densities $\Omega_bh^2$ and $\Omega_ch^2$.
The bounds on $w_0$ and the derived parameter $\Omega_m$ are obtained combining the performances of the 
surveys used in~\cite{Conley:2011ku} (low-z, SDSS, SNLS, HST) and the results are compared with the same paper.\\
What is shown in Figure~\ref{fig:sn} and Table~\ref{tab:sn} is that the obtained bound on the parameters agree 
with the results of~\cite{Conley:2011ku}, although the contour plot of $\Omega_m$ and $w_0$ can't reproduce the
non-Gaussian behavior of the actual posterior.

\begin{table}[h!]
\begin{center}
\begin{tabular}{|c|c|c|}
\hline
\hline
Parameter          & Fiducial & $68\%$ c.l. bound \\
\hline
$\Omega_m$         & $0.2$    & $0.1$\\
$w_0$              & $-0.9$   & $0.3$\\        
\hline
\end{tabular}
\caption{$68\%$ confidence level bounds on cosmological parameters to be compared with the first row
of Table 6 of~\cite{Conley:2011ku}.}\label{tab:sn}
\end{center}
\end{table}
\newpage
\section{Weak Lensing Forecasts}
The validation package also contains Weak Lensing forecasted bounds, obtained using the optimistic and pessimistic specifications for a ground based Dark Energy Task Force Stage IV (DETFIV) experiment~\cite{Albrecht:2006um}. 
Results are compared with what is obtained in the Weak Lensing section of the DETF document~\cite{Albrecht:2006um}.
While the optimistic case is in good agreement with the DETF forecasts, the pessimistic case is less degraded in
CosmicFish results; this is easily explained by the fact that we do not include the same systematic effects as in DETF pessimistic forecasts.

\begin{table}[h!]
\begin{center}
\begin{tabular}{|c|c|c|c|}
\hline
\hline
Parameter          & Fiducial  & Pessimistic case & Optimistic case \\
\hline
$\Omega_\Lambda$   & $0.73$    & $0.006$          & $0.005$\\
$w_0$              & $-1.0$    & $0.05$           & $0.05$\\     
$w_a$              & $0.0$     & $0.2$            & $0.2$\\   
\hline
\end{tabular}
\caption{$68\%$ confidence level bounds on cosmological parameters Dark Energy parameters to be
compared respectively with WL-IVLST-p and WL-IVLST-p entries of the Table at page 77 of~\cite{Albrecht:2006um}.}\label{tab:WL}
\end{center}
\end{table}

%\newpage

%
\section{Galaxy Clustering Forecasts}
\begin{figure}[h!]
\begin{center}
\begin{tabular}{ccc}
\includegraphics[width=6cm]{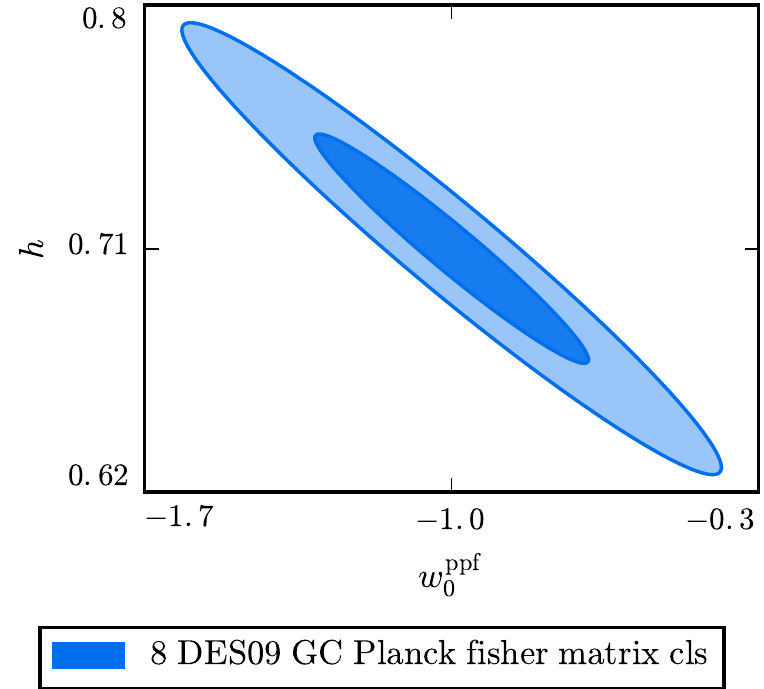}&
\includegraphics[width=6cm]{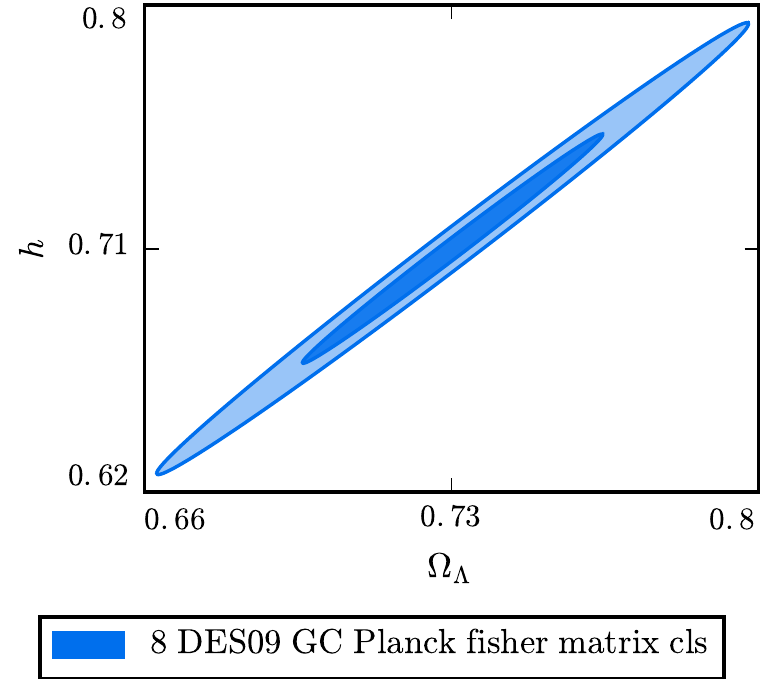}&
\includegraphics[width=6cm]{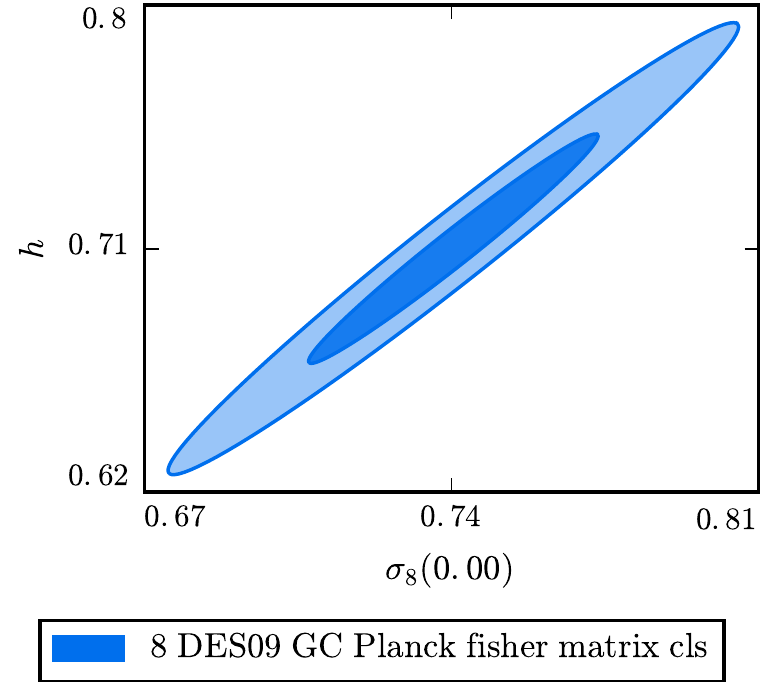}\\
\includegraphics[width=6cm]{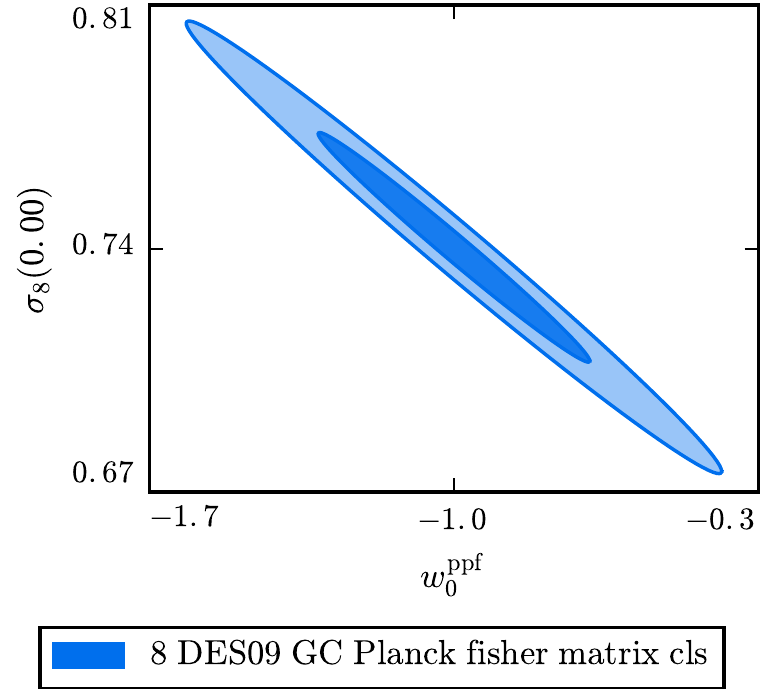}&
\includegraphics[width=6cm]{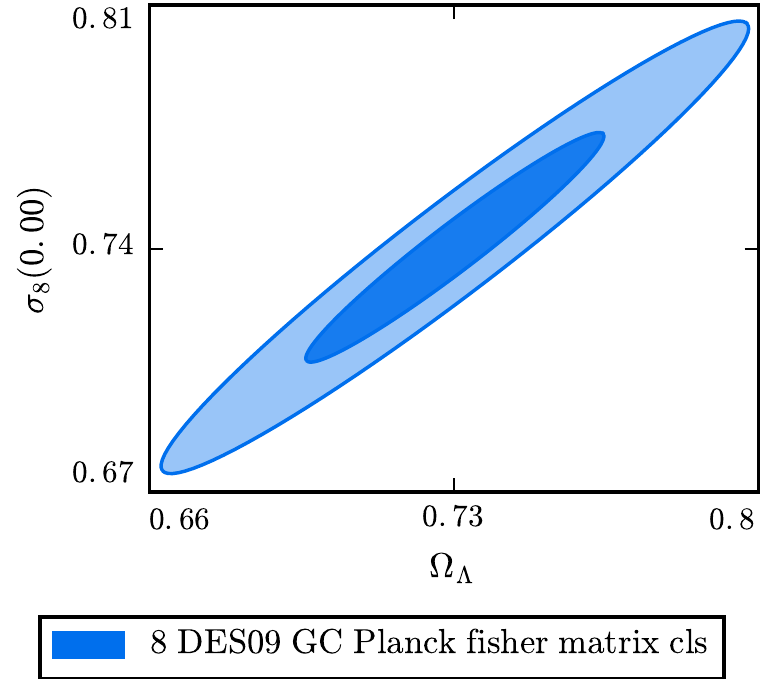}&
\includegraphics[width=6cm]{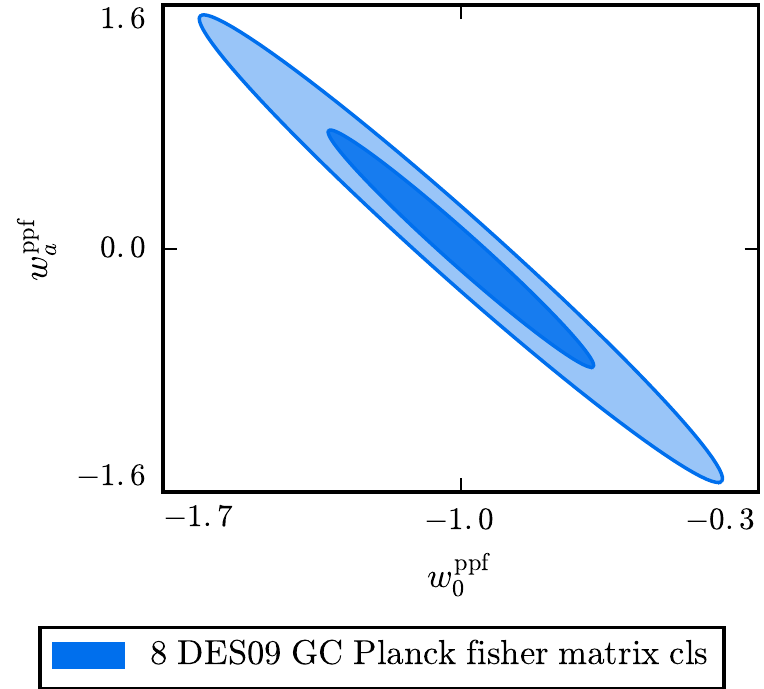}\\
\includegraphics[width=6cm]{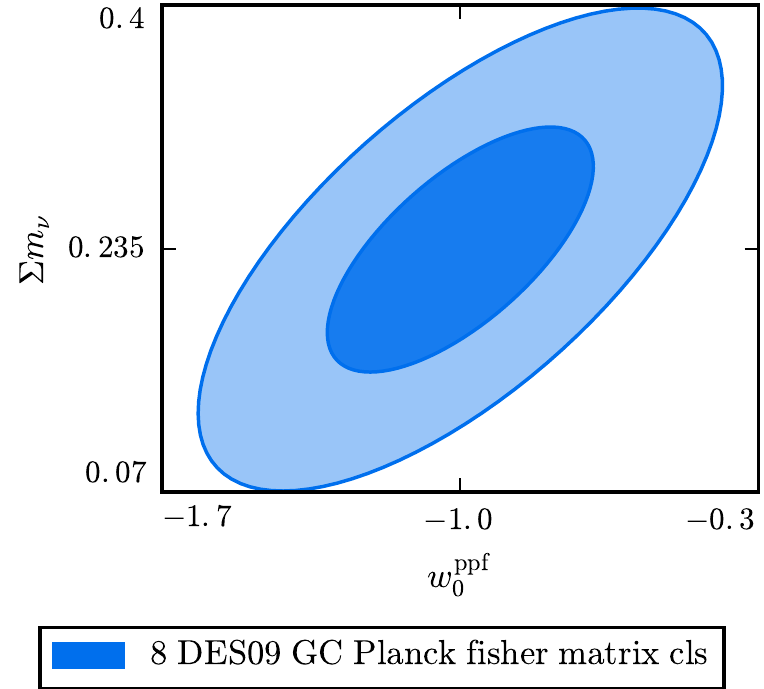}&
\includegraphics[width=6cm]{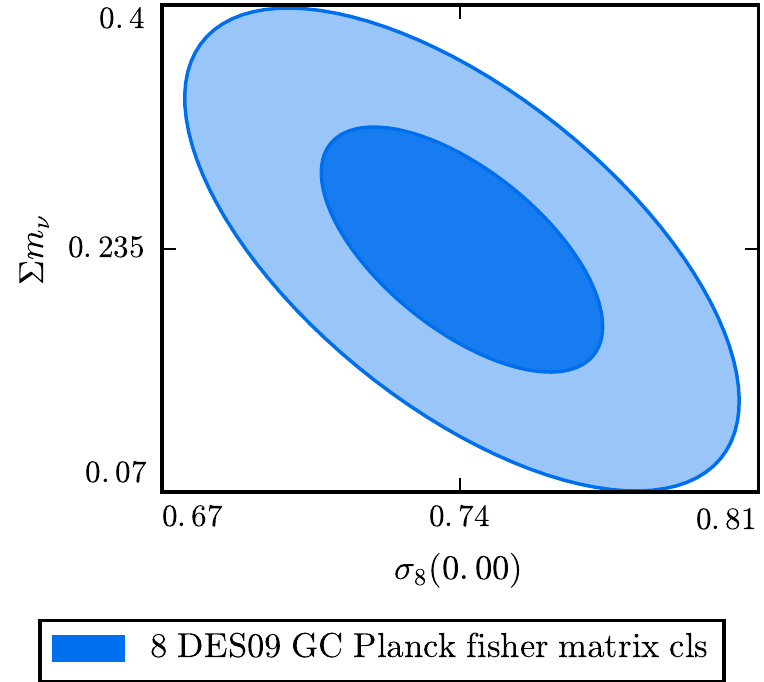}&
\includegraphics[width=6cm]{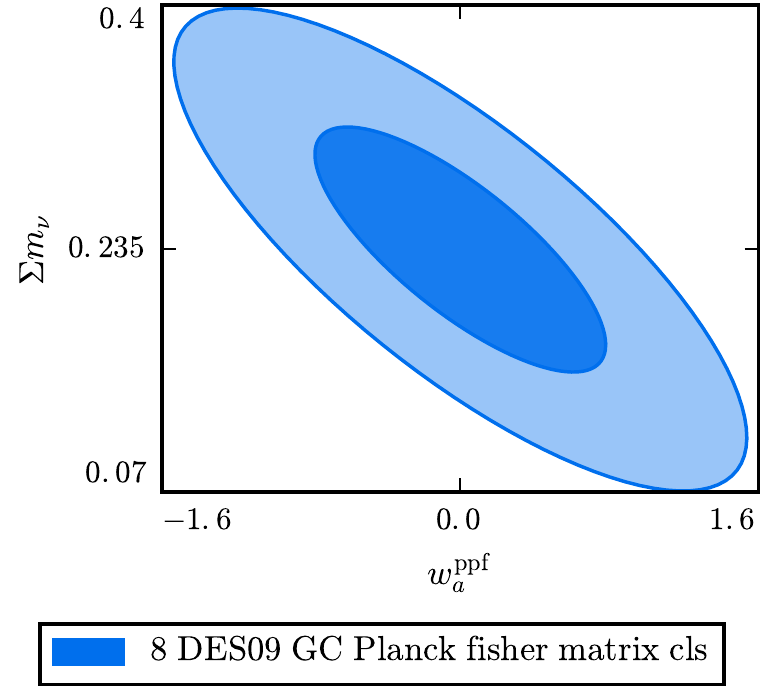}\\
\end{tabular}
\caption{Contour plots for several parameter combinations, to be compared with the blue dashed contours of 
Figure 11 of~\cite{Zablocki:2014ela}}\label{fig:gc}
\end{center}
\end{figure}

The CosmicFish Galaxy Clustering pipeline is validated obtaining bounds on cosmological parameters using DES 
specifications found in~\cite{Zablocki:2014ela}. As in this paper, the varying parameters are S6, the energy
density of massive neutrinos $\Omega_\nu h^2$ and the Dark Energy equation of state parameters $w_0$ and $w_a$.\\
Results shown in Figure~\ref{fig:gc} and Table~\ref{tab:gc} show a very good agreement with that is found in~\cite{Zablocki:2014ela}.

\begin{table}[h!]
\begin{center}
\begin{tabular}{|c|c|c|}
\hline
\hline
Parameter          & Fiducial & {\it Planck}+DES $68\%$ c.l. bound \\
\hline
$w_0$              & $-1.0$   & $0.3$\\
$w_a$              & $0.0$    & $0.8$\\
$\Sigma m_\nu$     & $0.32$   & $0.08$\\    
\hline
\end{tabular}
\caption{$68\%$ confidence level bounds on the sum of neutrino masses and on Dark Energy equation of states parameters.
To be compared with the third row, fifth column of Table V and fourth row, second column of Table VII of~\cite{Zablocki:2014ela}.}\label{tab:gc}
\end{center}
\end{table}

\acknowledgments

We are grateful to Ana Ach\'ucarro, Carlo Baccigalupi, Erminia Calabrese, Stefano Camera, Luigi Danese, Giulio Fabbian, Noemi Frusciante, Bin Hu, Valeria Pettorino, Levon Pogosian, Giuseppe Puglisi and Alessandra Silvestri for useful and helpful discussions on the subject. We are indebted to Luca Heltai for help with numerical algorithms.
MM is supported by the Foundation for Fundamental Research on Matter (FOM) and the Netherlands Organization for Scientific Research / Ministry of Science and Education (NWO/OCW). MM was also supported by the DFG TransRegio TRR33 grant on The Dark Universe during the preparation of this work.
MR acknowledges partial support by the Italian Space Agency through the ASI contracts Euclid-IC (I/031/10/0) and the INFN-INDARK initiative.
MR acknowledges the joint SISSA/ICTP Master in High Performance Computing for support during the development of this work.
MR thanks the National Astronomical Observatories, Chinese Academy of Science for the hospitality during the initial phases of development of this work.
MR and MM thank the Galileo Galilei Institute for Theoretical Physics for the hospitality and the INFN for partial support during the completion of this work.
GBZ and YW are supported by the Strategic Priority Research Program ``The Emergence of Cosmological Structures'' of the Chinese Academy of Sciences Grant No. XDB09000000, and by University of Portsmouth. YW is supported by the NSFC grant No. 11403034. 

%%%%%%%%%%%%%%%%%%%%%%%%%%%%%%%%%%%%%%%%%%%%%%%%%%%%%%%%%%%%%%%%%%%%%%%%%%%%%%%%%%%%%%%%%%%%%%%%%%%%%%%%%%%%

%\end{chapter}
\end{document}